**26/6/13**

# Understanding the effect resonant magnetic perturbations have on ELMs

A. Kirk[1], I.T. Chapman[1], T.E. Evans[2], C. Ham[1], J.R. Harrison[1], G. Huijsmans[4], Y. Liang[3], Y.Q. Liu[1], A. Loarte[4], W. Suttrop[5] A.J. Thornton[1]

[1]EURATOM/CCFE Fusion Association, Culham Science Centre, Abingdon, Oxon, OX14 3DB, UK
[2]General Atomics, P.O. Box 85608, San Diego, California 92186-5608, USA
[3]EURATOM-FZ Julich, D-52425 Julich, Germany
[4]ITER Organization, Cadarache, St. Paul-lez-Durance, France
[5]Max-Planck Institut für Plasmaphysik, EURATOM Association, Garching, Germany

**Abstract**

All current estimations of the energy released by type I ELMs indicate that, in order to ensure an adequate lifetime of the divertor targets on ITER, a mechanism is required to decrease the amount of energy released by an ELM, or to eliminate ELMs altogether. One such amelioration mechanism relies on perturbing the magnetic field in the edge plasma region, either leading to more frequent, smaller ELMs (ELM mitigation) or ELM suppression. This technique of Resonant Magnetic Perturbations (RMPs) has been employed to suppress type I ELMs at high collisionality/density on DIII-D, ASDEX Upgrade, KSTAR and JET and at low collisionality on DIII-D. At ITER-like collisionality the RMPs enhance the transport of particles or energy and keep the edge pressure gradient below the 2D linear ideal MHD critical value that would trigger an ELM, whereas at high collisionality/density the type I ELMs are replaced by small type II ELMs. Although ELM suppression only occurs within limitied operational ranges, ELM mitigation is much more easily achieved. The exact parameters that determine the onset of ELM suppression are unknown but in all cases the magnetic perturbations produce 3D distortions to the plasma and enhanced particle transport. The incorporation of these 3D effects in codes will be essential in order to make quantitative predictions for future devices.

## 1. Introduction

The reference scenario for ITER [1] is the high confinement mode (or H-mode) [2]. The improved confinement is the result of a transport barrier that forms at the plasma edge producing a steep pressure gradient region. This pressure gradient results in a repetitive plasma instability called an Edge-Localised Mode (ELM) [3][4]. Type I ELMs are explosive events, which can eject large amounts of energy and particles from the confined region [4]. Extrapolation from current measurements suggest that in order to ensure an adequate lifetime of the divertor targets on ITER the maximum ELM energy flux that can be repetitively deposited is 0.5 $MJm^{-2}$ [5]. Combined with assumptions on the ELM energy deposition profiles this leads to a maximum energy lost from the plasma during an ELM of $\Delta W_{ELM}$ = 0.66 MJ [6]. In the ITER baseline *Q*~10 (where *Q* is the fusion power gain factor = $P_{fusion}/P_{in}$) scenario, which has a plasma current ($I_P$) of 15 MA, the expected natural ELM frequency is ~ 1Hz with each ELM having $\Delta W_{ELM}$ ~ 20 MJ [7]. The ELM frequency scales as $f_{ELM} \propto I_P^{-1.8}$ [8], therefore, the natural ELM frequency in ITER will vary from ~ 1 Hz at for plasmas with $I_P$ = 15MA to $f_{ELM}$ ~7Hz for IP=5 MA (see Figure 1a). Assuming, as is observed on other devices (see [8] and reference therein), that the ELM size ($\Delta W_{ELM}$) multiplied by the ELM frequency ($f_{ELM}$) remains a constant fraction of the input power ($P_{in}$) (i.e. $\Delta W_{ELM} x f_{ELM}$ = 0.3-0.4x$P_{in}$) and taking into account the changes in the power deposition profile and the sharing between targets a mitigated ELM frequency required to keep the divertor energy flux density below 0.5 $MJm^{-2}$ can be calculated as a function of $I_P$ [9] and is shown in Figure 1a. For discharges with $I_P$>8MA some form of ELM mitigation (increase in ELM frequency over the natural value) is required. In addition to considering the effect of the ELMs on the PFCs, it has also been observed on ASDEX Upgrade that during operations with a tungsten (W) divertor a minimum ELM frequency is required to ensure that the W accumulation at the edge remains acceptable [10][11]. An analysis requiring that the edge W concentration remains below $2.5 \times 10^{-5}$ of the electron density [9] results in the minimum ELM frequency on ITER being ~ 18Hz. Combining the requirements on avoiding PFC damage and W accumulations results in the required increase in ELM frequency over the natural ELM frequency as a function of $I_P$ to be in the range ~3-40, which is shown in Figure 1b. However, the optimum solution in terms of divertor lifetime would in fact be complete ELM suppression, as long as it is accompanied by sufficient particle transport in order to avoid W accumulation.

Several ELM control techniques have been investigated for ITER (see [12] and references therein). These include pellet pacing, vertical kicks and the application of magnetic perturbations, which is the subject of this paper. This technique relies on perturbing the magnetic field in the edge plasma region, either leading to more frequent smaller ELMs (ELM mitigation) or ELM suppression. This technique of Resonant Magnetic Perturbations (RMPs) has been employed to either mitigate or suppress type I ELMs on DIII-D [13][16], JET [17], MAST [18], ASDEX Upgrade [19] and KSTAR [20]. Based on these results, a set of in-vessel coils is being considered as one of the two main systems for ELM control in ITER [5].

In this paper, the results from these devices and the possible way that the RMPs are affecting type I ELMs will be discussed. Section 2 describes the access conditions that have been found to be required to achieve type I ELM suppression. In section 3 what happens to the edge plasma during ELM suppression/mitigation will be explored. Section 4 describes the modelling that has been performed to try to describe the results obtained and section 5 give a summary and proposes possible mechanism that could explain the observations.

## *2. Access conditions for type I ELM suppression*

Resonant Magnetic Perturbations (RMPs) have been employed to mitigate and suppress type I ELMs at high collisionality on DIII-D [13], ASDEX Upgrade [19], KSTAR [20] and JET when operated with an ITER Like Wall (ILW) [21] and at low collisionality on DIII-D [14][15][16]. Figure 2a shows the regions of operational space for which type I ELMs have been suppressed in terms of pedestal collisionality ($\nu^*_e$) versus line average density expressed as a fraction of the Greenwald number ($n_e/n_{GW}$).

In addition to complete suppression of the type I ELMs, all of these devices can achieve periods of ELM mitigation. Similar periods of ELM mitigation have also been obtained on JET with a carbon wall [17][22] and MAST [18]. As can be seen in Figure 2b the region over which ELM mitigation has been achieved is wider than that for type I ELM suppression. In this section the experimental access conditions for ELM suppression will be discussed and in particular how the "islands" for suppression shown in Figure 2a come about.

### 2.1 Type I ELM suppression at low collisionality

DIII-D has demonstrated suppression of type I ELMs in a plasma with a similar shape and edge collisionality to the Q=10 ITER baseline scenario [16]. An example of such a discharge is shown in Figure 3 in which the RMPs are applied using two rows of internal off mid-plane coils (I-coils) in a n=3 even parity configuration (n being the toroidal mode number). Above a certain current in the I-coils type I ELMs are suppressed as long as the edge collisionality is below a threshold level ($\nu^*_e < 0.35$), if the collisionality is increased above this level then mitigated type I ELMs return (i.e. the ELM frequency is higher than in a similar shot with no RMPs) [15]. Full suppression is achieved in a limited window of the edge safety parameter ($q_{95}$). For the ITER similar shape discharges ELM suppression is typically found for an even parity configuration of the coils at $q_{95} = 3.5 \pm (0.05\text{-}0.225)$, where the size of the resonance window depends on the strength of the perturbation [23]. Outside the resonant window strong ELM mitigation is observed. ELM suppression was also achieved using a single row of the I-coils, but this required higher current per coil [24]. However, suppression was not obtained with a n = 3 perturbation of similar strength at the $q_{95}$ surface from a single-row of external, large aperture coils on the outer equatorial mid-plane [24].

In spite of several attempts type I ELM suppression at low $\nu^*$ has not been achieved on any other device. At low collisionality on ASDEX Upgrade ELM mitigation is observed but not full suppression [25]. On JET, the type-I ELM frequency in low collisionality ($\nu^*e \sim 0.1$) H-mode plasmas has been increased by a factor of up to 5 when applying static $n = 1$ or 2 fields produced by four external mid-plane error field correction coils (EFCCs) [17][22].

## 2.2 Type I ELM suppression at high collisionality/density

The first sustained suppression of type I ELMs in a high collisionality discharge was achieved on DIII-D [13]. Unlike in the low collisionality regime, here the coil configuration was n=3 odd parity which was not well aligned with the underlying resonant field resulting in a magnetic field perturbation at the q=11/3 surface an order of magnitude smaller ($\delta b_r/B_T \sim 1.6 \times 10^{-5}$ compared to $2.6 \times 10^{-4}$) than that required for ELM suppression in a low collisionality discharge [26]. Application of the coils in a resonant configuration (even parity) had little effect on the ELMs. A $q_{95}$ scan, performed by ramping the plasma current, revealed that type I ELM suppression only occurred in a limited window around $q_{95}$=3.7 [14]. Discharges were repeated at a similar plasma density (~$7.3 \times 10^{19}$ $m^{-3}$) but at a lower collisionality and it was found that type I ELMs, although mitigated, returned when $\nu^*_e < 2.0$ [15][14]. This suggests that there is a lower collisionality limit for suppression.

Suppression of type I ELMs has also been established at high collisionality in ASDEX Upgrade [19] using an internal off mid-plane coil set (called B-coils). An example is shown in Figure 4, where type I ELMs are suppressed as the plasma density crosses a certain threshold. While the suppressed ELM state has many similarities to that observed in DIII-D the access conditions are subtly different. For example, on ASDEX Upgrade suppression of type I ELMs can be obtained with n=2 magnetic perturbations that are resonant (odd parity) and not resonant (even parity) with the edge safety factor profile. Interestingly the required current in the B-coils is similar for the two coil configurations, in spite of the fact that the resonant field amplitude is a factor of 5.5 times higher in the odd parity case [19]. Type I ELM suppression also occurs over a wide range in $q_{95}$. One final difference between the access conditions on ASDEX Upgrade compared to DIII-D is that on ASDEX Upgrade the suppression of type I ELMs is associated with a plasma density expressed as a fraction of the Greenwald density ($n_{GW}$) rather than collisionality, with suppression being observed for $n_e/n_{GW}>0.53$ [27].

Type I ELMs have also been suppressed on KSTAR, this time using a resonant n=1 perturbation [20]. On JET, with a carbon first wall, no effect was observed at high collisionality [22], however, with the ILW suppression of type I ELMs with an n=2 perturbation has been achieved [21].

On MAST, while clear ELM mitigation has been observed, ELM suppression has not been established despite a large overlap in terms of the $\nu_e^*$ versus $n_e/n_{GW}>0.53$ operational space of the other devices (see Figure 2b) [28]. Hence just accessing the correct operational space in terms of density and/or collisionality is not sufficient to ensure type I ELM suppression.

## 3. *What happens during type I ELM suppression/mitigation*

### 3.1 Changes to the pedestal during type I ELM suppression at high collisionality/density

On both DIII-D [14] and ASDEX Upgrade [19] the pedestal density and temperature profiles remain largely unchanged in going from the type I ELM-ing to ELM suppressed periods. On ASDEX Upgrade there is evidence for a slight increase in particle confinement as the type I ELMs are suppressed [27]. The application of the magnetic perturbations has very little effect on the plasma rotation or the radial electric field in ASDEX Upgrade while in DIII-D the rotation is strongly damped and the radial electric fields is very weakly altered.

DIII-D [14], ASDEX Upgrade [19] and JET [21] have reported that the type I ELMs are replaced by small-scale high-frequency edge perturbations, which are reminiscent of type II ELMs. These bursts lead to an increase in the particle transport from the plasma [26] [29], which probably keeps the pedestal just below the level required to trigger a type I ELM.

**3.2 Changes to the pedestal during type I ELM suppression at low collisionality**

In low collisionality discharges in DIII-D after the I-coil current is turned on to create the RMP, the pedestal density and pressure drop [15]. The major part of the change in the pedestal pressure is due to the reduction in pedestal density. The maximum electron temperature gradient actually increases slightly [30]. There are significant changes in the edge toroidal rotation and the edge radial electric field when the I-coil is applied and the edge toroidal rotation of the carbon ions actually increases in the steep gradient region near to the separatrix [15].

The drop in edge pressure gradient moves the plasma into a region stable to peeling-ballooning modes [15], consistent with the suppression of type I ELMs. The question is why the pedestal in the ELM suppressed discharge stops evolving i.e. why it does not reach the peeling ballooning boundary. Possibly this is because the cross field particle transport in the presence of the RMPs is larger than the particle transport averaged over ELMs. This enhanced particle transport occurs over a wider range of $q_{95}$ compared to the ELM suppression window. The density fluctuations during the ELM suppressed stage loose their bursty character (i.e. they differ from the type II ELM-like fluctuations observed at high collisionality) but the base level rises [26].

The effective particle confinement time, estimated from the pellet retention time, is found to be reduced during ELM suppression compared the ELM-ing discharge [16]. In addition the ion scale fluctuations are observed to increase across a wide region of the plasma [31]. The fluctuations are observed to increase where the ExB shearing rate is reduced, which is possibly a result of the RMP induced torques on the plasma.

**3.3 Pedestal evolution during the mitigated stage**

Although the full suppression of type I has not been achieved in MAST, both JET (with the carbon wall) [22] and MAST [28] have demonstrated a reduction of the ELM energy losses as the ELM frequency increases compatible with $\Delta W_{ELM} x f_{ELM} \sim$constant. The ELM frequency was increased by a factor of 4 on JET and 9 on MAST. An example of the inter-ELM pedestal evolution observed in MAST for natural and mitigated ELMs is shown in Figure 5a. The pedestal electron density and temperature evolve in a similar way between ELMs in the shots with and without RMPs; however, in the shots with RMPs applied the ELM is triggered earlier in the cycle at a lower value of $P_e^{ped}$, reflecting the increased ELM frequency.

Stability analyses have been performed for both the JET [32] and MAST [28] mitigated discharges using the ELITE stability code [33] assuming toroidal symmetry. In both cases the analysis shows that the experimental point moves from the peeling ballooning boundary (a trait often associated with type I ELMs) for the natural ELM to point which is significantly below the threshold. Hence it would appear that the ELMs are being triggered in a region that was previously stable to peeling-ballooning modes.

One possible way to explain this would be that the mitigated ELMs are not type I. To investigate this on JET the ELM frequency has been measured as a function of input beam power ($P_{NBI}$). The ELM frequency is found to increase with $P_{NBI}$ for both mitigated and unmitigated ELMs, consistent with the mitigated ELMs still being type I ELMs just smaller in size and at a higher frequency [22]. On MAST a comparison of the filament structures observed during the ELMs in the natural and mitigated stages show that they both have similar characteristics [34]. Based on the toroidal mode number of the filaments it would appear that the mitigated ELMs still have all the characteristics of type I ELMs even though their frequency is higher, their energy loss is reduced and the pedestal pressure gradient is decreased.

While the mitigated ELMs are smaller there is a price to be paid in terms of plasma confinement. Since the ELM is triggered earlier in the pedestal pressure evolution, the peak and average pedestal pressure is reduced in the ELM mitigated shots. Due to the stiffness of the profiles this also leads to a reduction in the overall stored energy in the plasma. Figure 5b shows a plot of the plasma stored energy in the mitigated shots on MAST as a fraction of the stored energy in the shot without RMPs applied as a function of the increase in ELM frequency. There is an initial sharp drop in confinement which then levels out at between a 15-20 % loss of stored energy at the highest ELM frequencies.

### 3.4 Changes to the plasma shape

The application of non-axisymmetric perturbations fields have been observed to produce a toroidally varying displacement to the mid-plane outer radius of the plasma on DIII-D [35][36], ASDEX Upgrade [27], MAST [37] and JET [38]. In some devices the displacement is smaller than on others. The displacement of the edge of the plasma scales approximately linearly with the maximum resonant component of the applied radial field ($B_r^{res}$) (see Figure 6). The applied fields have a significant effect on the toroidal periodicity of the plasma edge, deforming the separatrix by a few per cent of the minor radius.

Whilst two-dimensional treatment of the plasma equilibrium is routine, three-dimensional plasma equilibrium reconstruction is more difficult. On MAST, when the n = 3 RMPs are applied to control ELMs, the displacement of the plasma boundary, which is measured at various toroidal locations, varies toroidally by up to 5% of the minor radius. The empirically observed corrugation of the plasma edge position agrees well with three-dimensional ideal plasma equilibrium reconstruction using the VMEC code [39]. The influence of the 3D corrugation on infinite-n ballooning stability has been examined using the COBRA code [40]. The growth rate of the $n=\infty$ ballooning modes at the most unstable toroidal location is a factor of two larger than the axisymmetric case i.e. the plasma edge is strongly destabilised at certain toroidal positions [41].

In an ideal axi-symmetric poloidally diverted tokamak the magnetic separatrix (or last closed flux surface) separates the region of confined and open field lines. The idea that so-called separatrix “manifold” structures could exist was first introduced to the tokamak community by Roeder, et al., [42] and Evans et al., [43][44]. Non-axi-symmetric magnetic perturbations split this magnetic separatrix into a pair of so called "stable and unstable manifolds" [43][45]. Structures are formed where the manifolds intersect and these are particularly complex near to the X-point. The manifolds form lobes that are stretched radially both outwards and inwards. Some of these lobes can intersect the divertor target and result in the strike point splitting often observed during RMP experiments [46][47].

On MAST, these lobes structures have been observed using filtered visible imaging (Figure 7a) [48]. A clear correlation is observed between the size of the lobe length and the change in ELM frequency [28], which may suggest that the lobes themselves are having a direct impact on the stability of the edge plasma to peeling ballooning modes. On DIII-D, similar structures have been measured by tangential imaging of extreme ultra-violet and soft x-ray emission (Figure 7b) [49]. The comparison of the lobes structures with modelling will be discussed in the next section.

## *4. Modelling the effect of RMPs*

The original interpretation of the low collisionality discharges on DIII-D was that the RMPs would create magnetic islands at the rational surfaces at the plasma edge and the overlap of these neighbouring islands would create an ergodised edge region. This ergodic layer at the edge of the plasma would enhance particle and heat transport, which would reduce the pedestal pressure and hence avoid the triggering of a type I ELM [16].

Vacuum magnetic modelling (in which the magnetic field generated by the plasma in response to the RMPs is neglected) indicates that the n=3 perturbations from the DIII-D I-coils create a set of magnetic islands centred on the resonant surfaces, which overlap at the edge, resulting in magnetic stochasticity [16].

The Chirikov parameter ($\sigma_{\text{Chirikov}}$) is a measure of the island overlap and assuming that the RMPs have a single n component, this parameter is defined as $\sigma_{Ch} \equiv (\delta_m + \delta_{m+1})/\Delta_{m,m+1}$, where $\delta_m$ and $\delta_{m+1}$ represent the half-widths of the magnetic islands on the q=m/n and q=(m+1)/n surfaces (m being the poloidal mode number and q the safety factor) and $\Delta_{m,m+1}$ the distance between these two surfaces.

A study on DIII-D [23] showed that ELM suppression is well correlated with the condition $(\Delta\psi_{pol})_{\sigma_{Ch}>1} > 0.17$, where $(\Delta\psi_{pol})_{\sigma_{Ch}>1}$ is the width, in terms of normalised poloidal flux $\psi_{pol}$, of the region where $\sigma_{Ch} > 1$ (i.e. the stochastic region). The criterion $(\Delta\psi_{pol})_{\sigma_{Ch}>1} > 0.17$ was used in order to calculate the current requirements in an early ITER ELM control coils design study [50] as well as for the latest ITER ELM coil design [51]. The maximum current capability of the ITER ELM control system based on this criteria, allowing for an additional 20 % margin, is 90 kAt. Including other harmonics may mean that the ITER coils will be able to produce an even larger level of ergodisation and hence have an even larger margin for success [52].

As discussed in section 3.2, in the low collisionality ELM suppressed discharges in DIII-D, the RMPs cause a decrease in the plasma density but have little effect on the electron temperature. This seems quite paradoxical, since it would be expected that an ergodic region, if created, should mainly enhance electron heat transport [53], which would reduce the electron temperature and temperature gradient.

However, the Chirikov parameter should be calculated taking into account the plasma response, since in order to stochastise the magnetic field at the edge, the field first needs to penetrate into the plasma. It is known that this tends to be prevented by the plasma rotation (see [54] and reference therein). Experiments [55][56], theory [57][58] and modelling [59] show that there is a threshold in terms of RMPs amplitude above which the rotation is stopped at the resonant surfaces and penetration can occur. When this happens, the plasma may amplify the RMPs. In order to fully model the physics of the pedestal region a full nonlinear two-fluid MHD model is most likely required [60][61][62][63]. It has been suggested [64][65] that the RMPs are screened due to the perpendicular rotation of the electron fluid and that for a low resistivity (or ideal) plasma the RMP field will only be large close to rational surfaces where the total electron perpendicular velocity ($V_e^{\perp}$) is near zero. As was discussed in references [63] and [36] one possible reason why the RMP field may be less screened in this region is because the resistivity is large and hence the screening currents are reduced. These effects are observed in the two fluid MHD modelling [62][63].

The measurements of the lobes structures observed near to the X-point on MAST and DIII-D can be used to estimate the penetration of the field. On MAST vacuum modelling gives a good quantitative agreement between the number and separation of the lobes, however, there appears to be a discrepancy in their radial extent [28], with the experimental lobes being shorter. Calculations have been performed using the MARS-F code, which is a linear single fluid resistive MHD code that combines the plasma response with the vacuum perturbations, including screening effects due to toroidal rotation [66] and realistic values of resistivity. The resistive plasma response significantly reduces the field amplitude near rational surfaces and reduces the resonant component of the field by more than an order of magnitude. The lobe extent predicted from the screened field is found to be in good agreement with the experimental observations [28].

On DIII-D the linear two fluid M3D-C1 code shows that the applied field is well screened except where $V_e^{\perp} \sim 0$ and here the field is in fact amplified [49][67]. As will be discussed in the next section it is thought that the amplification in this region may induce an island at the top of the pedestal and the transport due this island impedes the widening of the pedestal, which stops the peeling ballooning limit being reached [67].

## *5. Summary and possible mechanisms for ELM suppression/mitigation*

In order to avoid unacceptable erosion and possible damage to plasma facing components in ITER a method of reducing the size of type I ELMs is required. Several techniques have been considered. The only method that has been found to completely suppress type I ELMs is the application of resonant magnetic perturbations. To date complete suppression has been established either below a certain value of collisionality or above a collisionality/density threshold. In between these two limits ELM mitigation is observed.

At high ν*/density, there is little dependence on strength of perturbation or alignment but a clear high density(AUG)/ν* (DIII-D) threshold exists. There is little sensitivity on the rotation profile and the magnetic perturbations have little effect on the density or temperature profiles. At low ν*, there is a strong dependence on the strength and alignment ($q_{95}$) of the perturbation. There appears to be a clear ν* threshold ($\nu^* < 0.35$) and a strong dependence on the edge rotation.

Before discussing possible explanations for these two suppression windows, it is interesting to note that it is possible to replace type I ELMs by small/no ELM regimes without the application of RMPs (see [68] and references therein). These small/no ELM regimes show considerable reduction of instantaneous ELM heat load and their operational space can be characterised, in a similar way to the RMP type I suppressed regimes, in a plot of collisionality versus plasma density as a function of the Greenwald density (see Figure 8). For example, QH-mode and type II ELMs can be produced in discharges with good confinement properties at low and high collisionality respectively. At present these schemes are not considered in the ELM control schemes for ITER since they require operational parameters (shaping, rotation profiles etc) that are not compatible with ITER baseline operation.

Type II ELMs are typically observed in strongly shaped plasmas in a quasi double null (DN) magnetic configuration at high density [69][70]. For similar plasma conditions (density, heating power etc.) type II ELMs replace type I ELMs as the plasma approaches a DN but the changes in shape at the transition from type I to type II ELMs are only moderate [71]. For a fixed plasma shape there is a clear density ($\nu^*$) access threshold. Associated with a stable type II regime is a large radial flux of particles, where the particle flux is larger than that in a type I regime [72]. In fact, if the particle flux is not sufficient a mixed type I/II regime results [71] presumably because the pedestal continues to evolve until it reaches the peeling-ballooning boundary and results in a type I ELM. The mode responsible for type II ELMs appears to be driven unstable by changes in shape, density and possibly the change in neutral fuelling location as the second strike point becomes more important and has a higher toroidal mode number and very regular mode structure [71].

QH-mode is obtained at low density (collisionality) with a large edge velocity shear [73]. In such a configuration an edge harmonic oscillations (EHO), thought to be due to a saturated kink-peeling mode, is present and gives rise to enhance particle transport across the pedestal and into the scrape off layer (SOL). The existence of a strong EHO is essential, but a theory to explain this enhanced transport is still required. The maximum in the density fluctuations due to the EHO is located at the pedestal top. QH-mode plasmas operate near but below the peeling-ballooning stability boundary, with the enhanced transport due to the EHO stopping the pedestal evolving towards the boundary.

**5.1 Possible explanation for type I ELM suppression at high collisionality/density**

In type I ELM suppression at high collisionality/density using magnetic perturbations the type I ELMs are replaced by small edge fluctuations that have all the hallmarks of type II ELMs. It is know that type II ELMs are driven unstable by changes in shape, density and fuelling. Therefore it seem likely that the magnetic perturbations produce changes in plasma shape or changes to the neutral particle fuelling that allow the mode responsible for type II ELMs to be excited in the plasma. Then provided the particle transport from this mode is sufficient the peeling ballooning boundary cannot be reached and type I ELMs are suppressed. If this is indeed the process then it is essential that we understand type II ELMs and in particular how the mode that is responsible for them is triggered and how the particle transport can be enhanced so as to avoid triggering type I ELMs.

## 5.2 Possible explanation of type I ELM mitigation

The results presented in this paper and elsewhere suggest that ELM mitigation due to RMPs results from the 3D perturbations to the separatrix, which then cause a degradation of the edge stability to peeling ballooning modes. As depicted in Figure 9, in a natural ELM cycle the pressure pedestal height and width increase until the peeling ballooning limit is reached. The application of the RMPs leads to a 3D corrugation of the mid-plane separatrix leading to a pressure gradient that is no longer axi-symmetric, which combined with the lobe structures near to the X-point, leads to a decrease in the stability boundary [41]. The inter-ELM transport appears to be the same in the natural and the mitigated ELMs meaning that the pressure profile reaches this new, lower stability limit earlier in the natural ELM cycle and hence an increase in ELM frequency results. The level of the ELM mitigation achieved would then depend on the location of this new stability limit. The price paid for the mitigated ELMs, however, is a reduction in the maximum pedestal height achieved and hence in the overall stored energy. As will be discussed below, in order to achieve ELM suppression a mechanism would then need to be found to stop the pedestal evolving towards the stability boundary.

## 5.3 Possible explanation of type I ELM suppression at low collisionality

Type I ELM suppression at low collisionality using RMPs results because the pressure gradient is below the peeling ballooning stability limit. The questions are how does it get there and where does the transport come from to stop it evolving back to the Peeling-Ballooning (PB) boundary? The main contribution to the pressure gradient decrease is the pedestal density drop—the so-called ‘pump-out effect’, while the pedestal temperature does not drop and might even increase. As discussed above, at a first glance this seems to be contrary to the idea that an ergodic layer has been created since then both the density and electron temperature should decrease together since both diffusion and electron heat conductivity coefficients should rise simultaneously. Fluid models have been used to describe the density pump out event [74][75][76]. In [76], the observed phenomena are explained as a result of changes in the ambipolar electric field, which is modified by the RMPs. The importance of the changes in the radial electric field on the particle transport have also been identified from modelling performed using the kinetic XGC0 code [77]. It is found that, due to the kinetic effects, the stochastic parallel thermal transport is significantly reduced compared to the prediction from the standard Rechester–Rosenbluth model [53]. The parallel electron heat transport is found to be approximately the same as the particle transport, which is significantly enhanced due to the changes in the radial electric field ($E_r$). The trapped particles experience a net toroidal drift due to the changes in $E_r$ while the passing particles do not.

While these results go some way to explaining the enhanced transport, in order to stop the pedestal evolving towards the Peeling Ballooning boundary and triggering a type I ELM, the enhanced transport needs to be located near to the top of the pre-ELM pedestal. Recent findings on DIII-D suggest that the RMPs may induce an island at the top of the pedestal and the transport due this island impedes the widening of the pedestal, which then stops the peeling ballooning limit being reached [67]. Two fluid MHD modelling suggest that the island is formed on q=10/3 surface, which is where the electron perpendicular velocity is zero [78].

Rather than requiring an island to form in this region, an alternative explanation could be that sufficient transport is produced across this region as is produced in the QH-mode by the EHO. For example, it has been suggested that in the vicinity of rational magnetic surfaces, the infinite-n ideal MHD ballooning stability boundary is strongly perturbed by the 3-D modulations of the local magnetic shear associated with the presence of near resonant Pfirsch-Schlüter currents [79]. The resulting Kinetic Ballooning modes that are destabilised may then provide the transport required. An alternative model describes the enhanced transport in terms of RMP-flutter-induced plasma transport, which is found to be large enough to reduce plasma gradients in pedestal top region [80].

In order to extrapolate ELM suppression at low collisionality to other devices it is essential that the mechanisms for stopping the evolution of the pedestal are identified. The optimum would be to arrange things in such a way that this was achieved with the minimum reduction in pedestal height and hence plasma performance and to do this we need to understand the particle transport mechanism.

**5.4 Implication for ITER**

Although the maximum current capability of the ITER ELM control coils was based on a vacuum island overlap criteria, which may be questionable due to questions of plasma screening, an alternative criteria does not as yet exist. Modelling carried out for ITER [63] shows that the plasma rotation gives a strong screening of central islands and limits the RMP penetration. However it does suggest penetration into the pedestal top region. While ELM suppression has been achieved in limited operational space windows, ELM mitigation, which after all is all that is required for ITER, has been established over a wide operational window. ELM mitigation appears to be related to the distortions to the 3D plasma shape and calculations for ITER show that mid-plane displacements comparable with current devices should be achieved (Figure 6). Hence ELM mitigation should well be achievable but the questions that remain are will the increase in ELM frequency be sufficient and what will be the price in terms of energy confinement.

In the case of ELM suppression at low collisionality there is also a price to be paid in confinement due to the density pump out that appears to be a required part of the suppression mechanism. The best that could be hoped for would be to arrange the transport such that the pedestal just remained below the Peeling-Ballooning boundary of the natural ELM. To quantify this we will need the development of 3D stability codes coupled with a better understanding of the transport physics in this region.

The optimum may be if a suppression regime, similar to that obtained at high collisionality/density where the type I ELMs were replaced by another small ELM regime with little change in the pedestal parameters, could be achieved at ITER. While it is unlikely that a type II ELM regime would be obtainable at ITER collisionalities it may be possible to trigger another small ELM regime (for example Grassy ELMs [68]) using the application of the 3D fields. The reason for the uncertainties is that although we know the parameter space in the absence of RMPs for these regimes we do not as yet understand the physical mechanism responsible for them. These questions should be addressed through experiments and modelling in the near future.

While the exact parameters that determine the onset of ELM suppression are unknown, in all cases the magnetic perturbations produce 3D distortions to the plasma and enhanced particle transport. The incorporation of these 3D effects in codes will be essential in order to make quantitative predictions for ITER.

## Acknowledgement

This work was part-funded by the RCUK Energy Programme [grant number EP/I501045] and the European Communities under the contract of Association between EURATOM and CCFE. To obtain further information on the data and models underlying this paper please contact PublicationsManager@ccfe.ac.uk. The views and opinions expressed herein do not necessarily reflect those of the European Commission or of the ITER Organization.

**Figures**

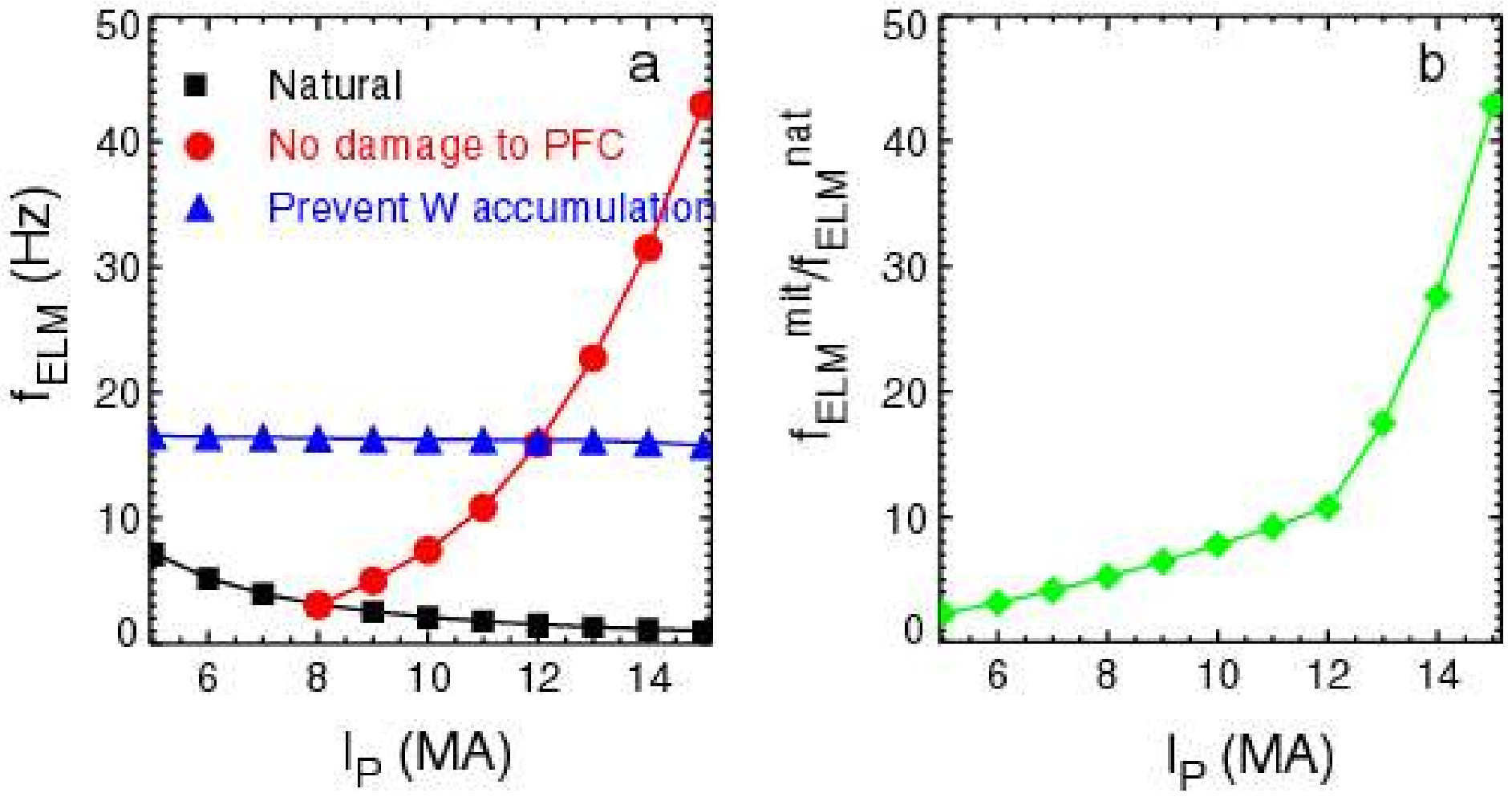

**Figure 1** a) Predicted ELM frequency ($f_{ELM}$) as a function of plasma current ($I_P$) for ITER for natural ELMs (squares), in order to avoid damage to plasma facing components (PFC) (circles) and to prevent W accumulation (triangles). b) The mitigated ELM frequency ($f_{ELM}^{mit}$) as a fraction of the natural ELM frequency ($f_{ELM}^{nat}$) as a function of $I_P$.

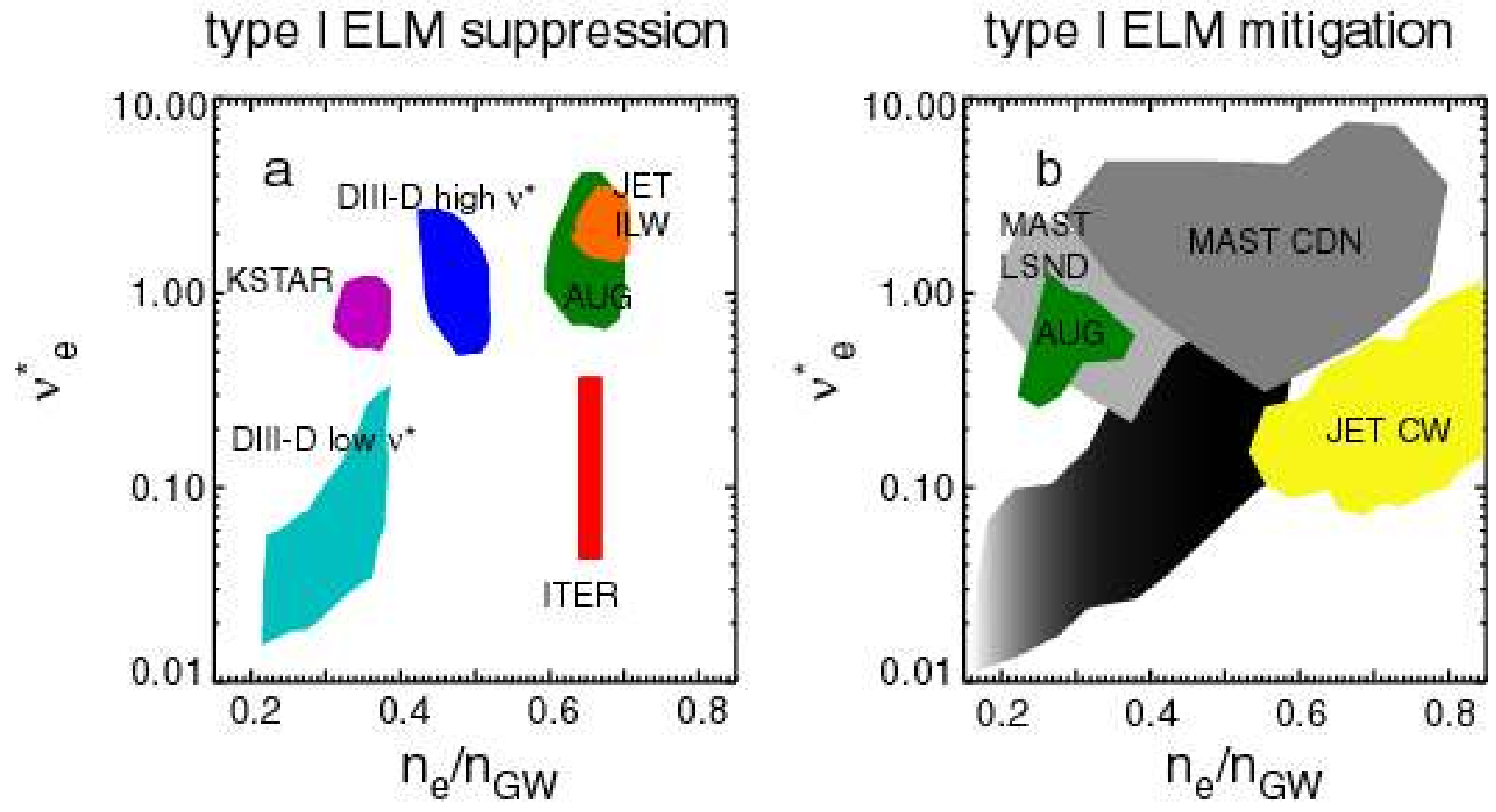

**Figure 2** Experimentally determined access condition in terms of pedestal collisionality ($\nu^*_e$) versus pedestal density as a fraction of the Greenwald density ($n_e/n_{GW}$) for a) suppression of type I ELMs and b) type I ELM mitigation.

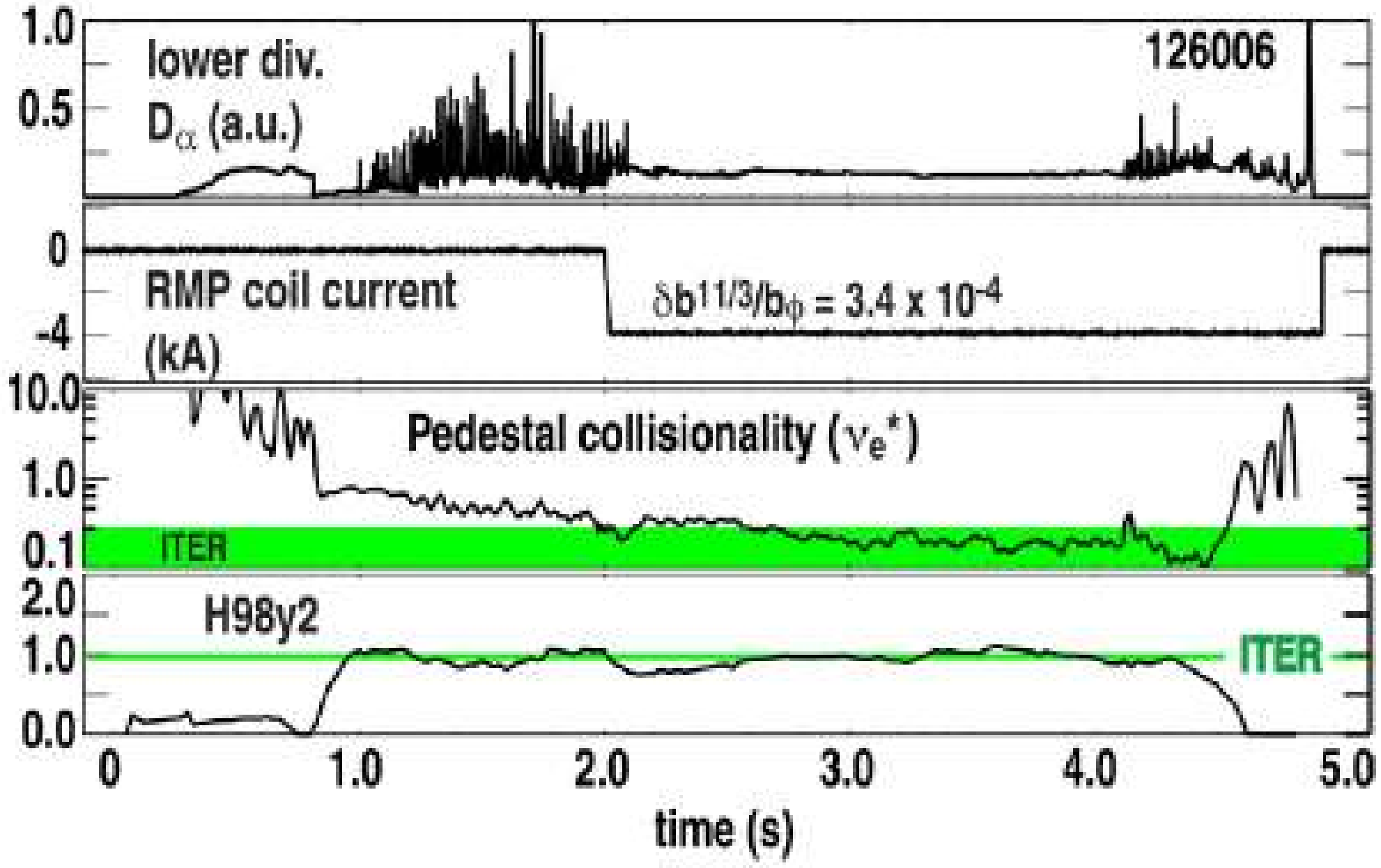

**Figure 3** Time traces of a) Lower divertor $D_\alpha$, b) Current in I-coil, c) electron pedestal collisionality and d) $H_{98y2}$ confinement scaling factor for a DIII-D discharge in which type I ELMs are suppressed. The values of pedestal collisionality and H98y2 expected by ITER are indicated in c) and d) respectively [16].

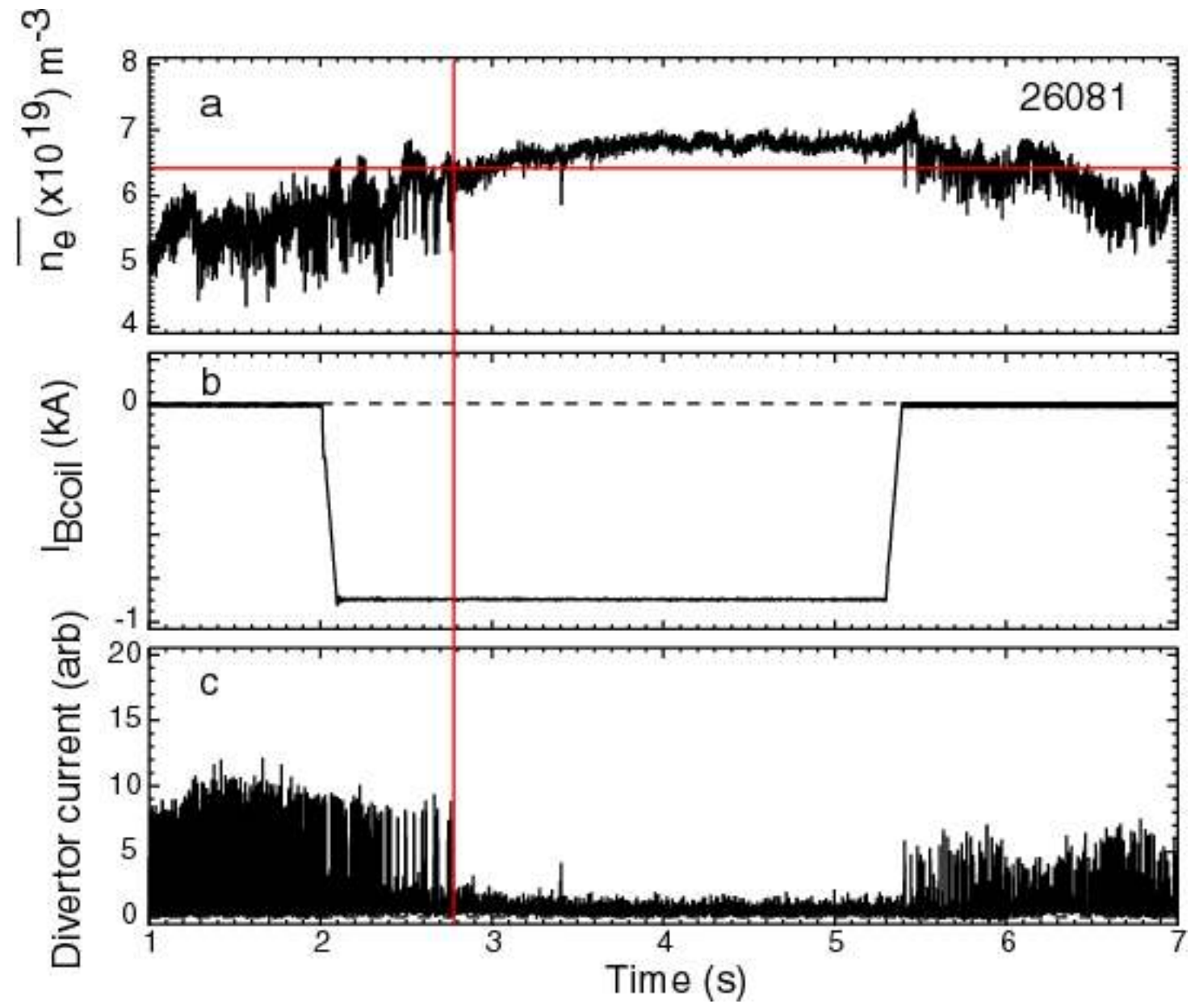

**Figure 4** Time traces of a) line averaged density, b) B-coil current and c) divertor current for a ASDEX Upgrade discharge in which type I ELMs are suppressed.

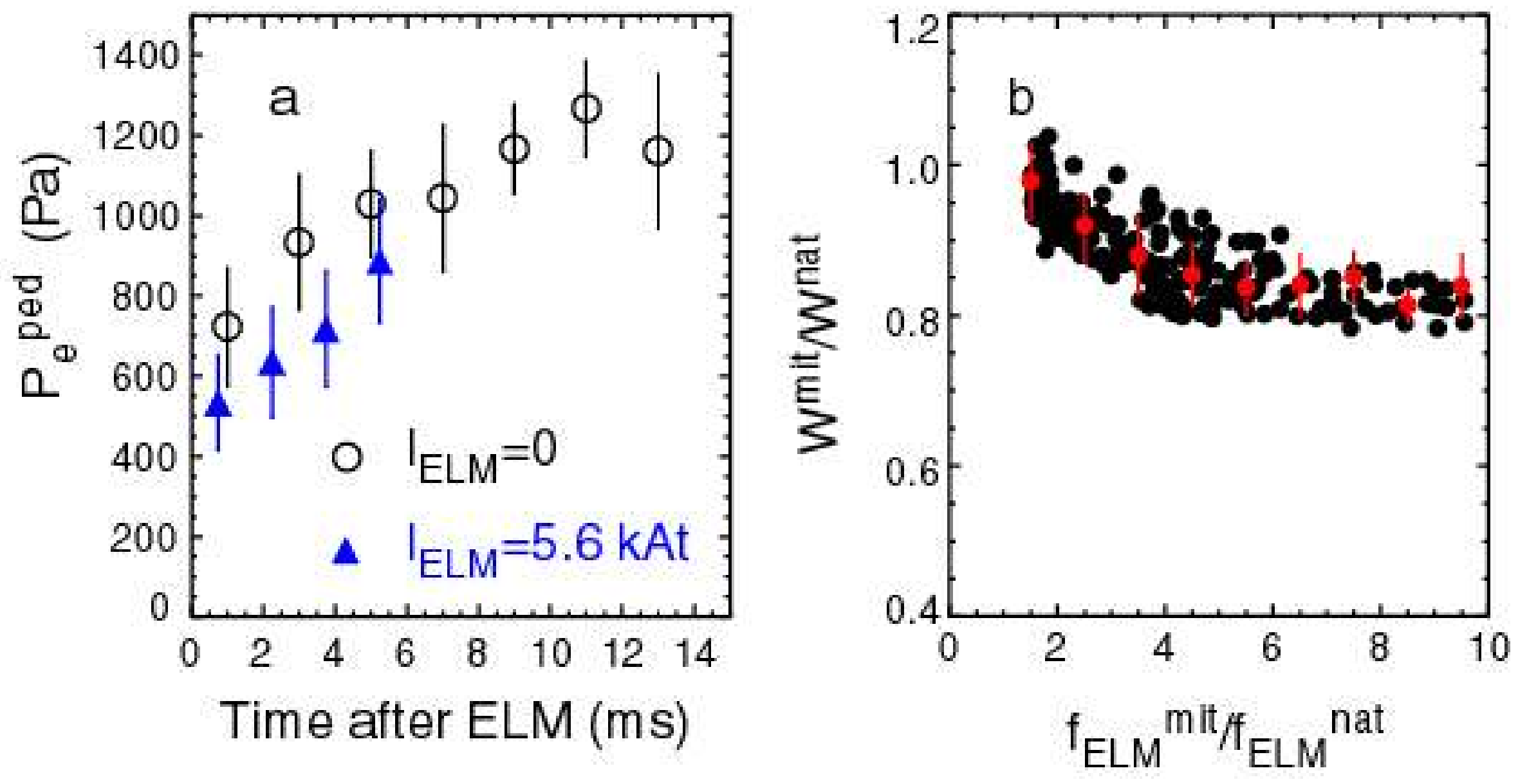

**Figure 5** a) Evolution of the electron pressure pedestal height during the ELM cycle for shots without (circles) and with (triangles) RMPs in an n=6 configuration on MAST. b) Plasma stored energy in the mitigated stage as a fraction of the stored energy in the shot without RMPs applied versus the fractional increase in ELM frequency (the points with error bars represent the binned mean and standard deviation of the underlying distribution).

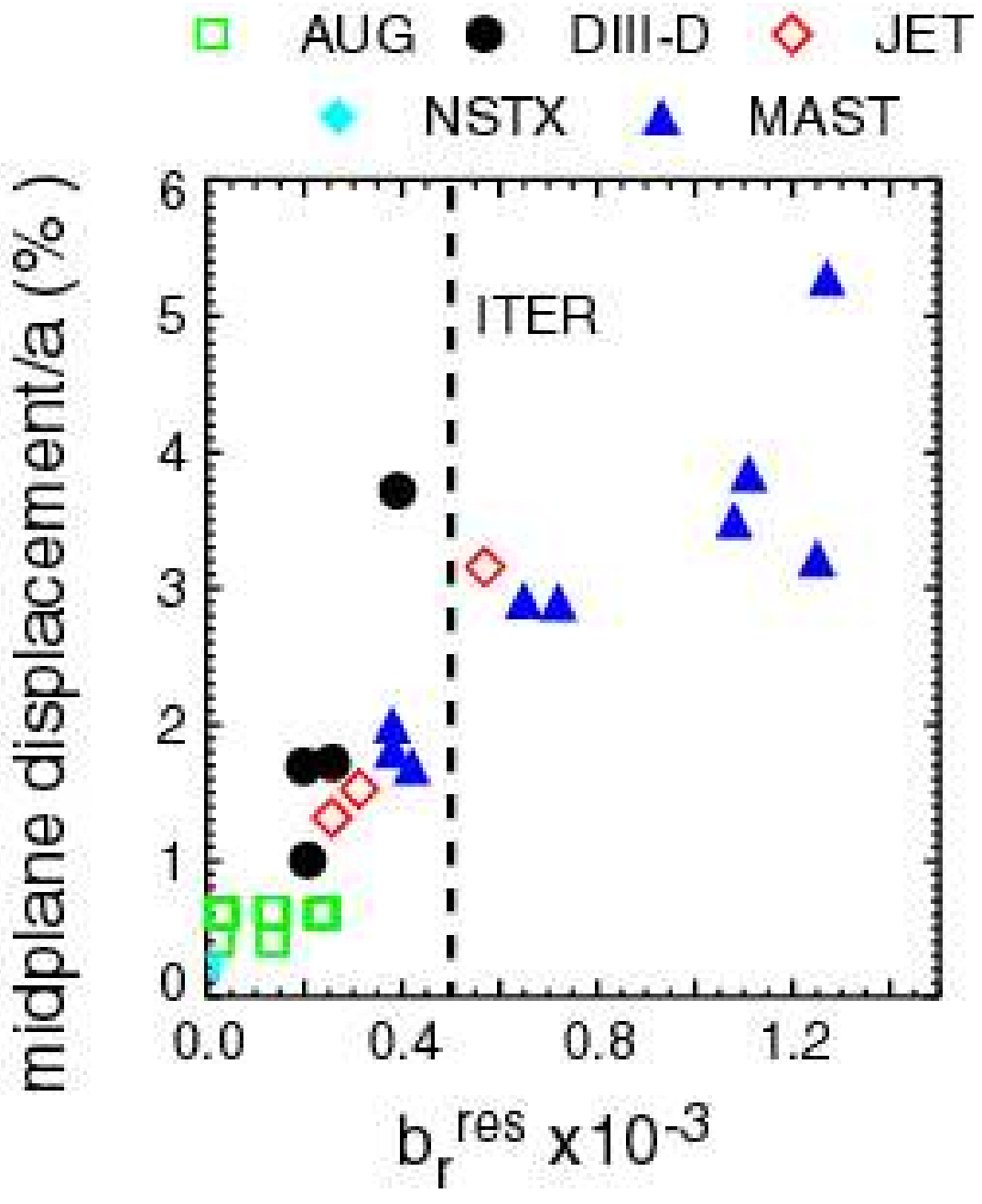

**Figure 6** Mid-plane displacement of the plasma expressed as a fraction of the minor radius as a function of the maximum resonant field component ($b_r^{res}$).

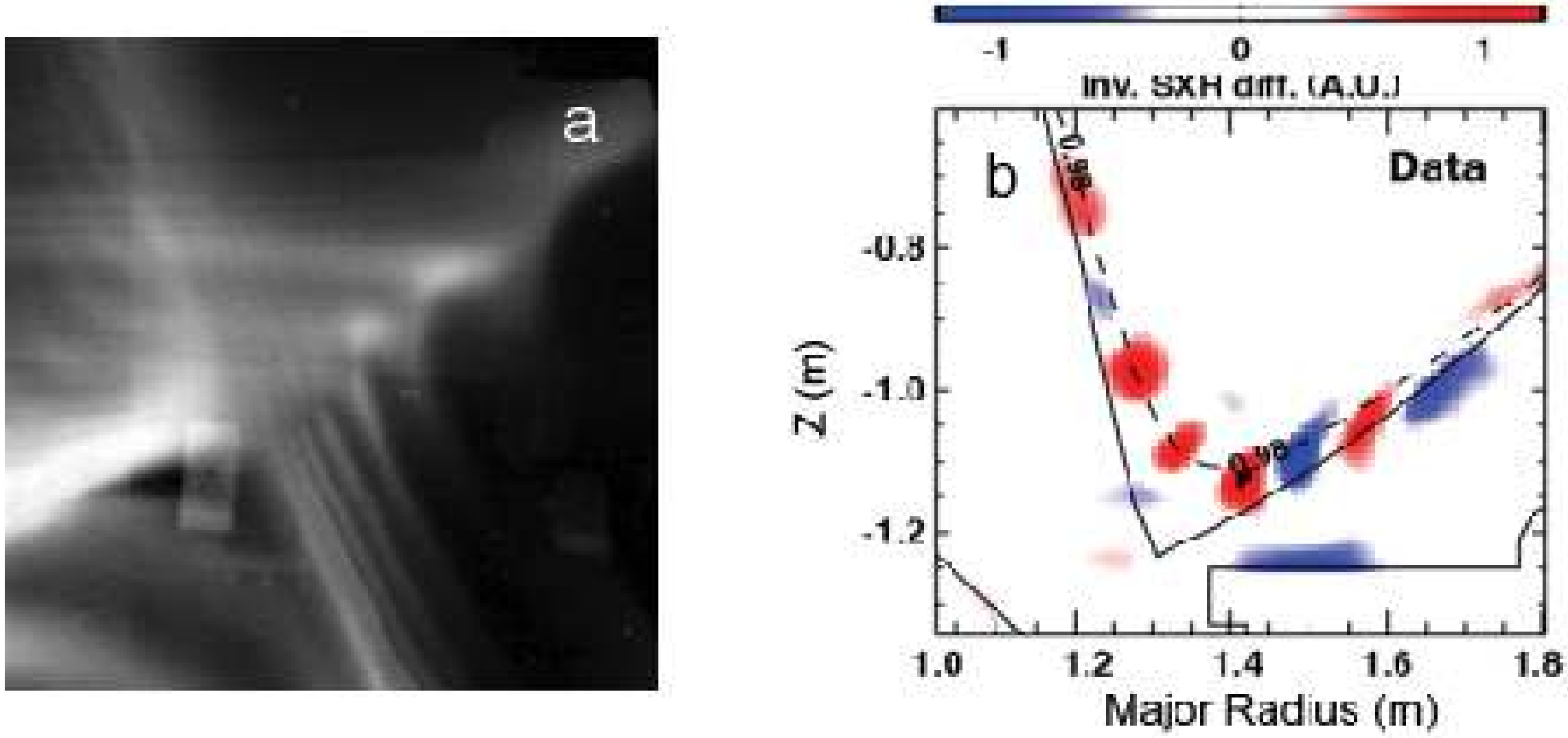

**Figure 7** a) Image of the $He^{1+}$ emission from the X-point region on MAST captured during an Inter-ELM period of a LSND H-mode with the RMPs in an n=6 configuration b) out of phase subtracted soft X-ray image if the X-point region on DIII-D during the application of RMPs in an n=3 configuration.

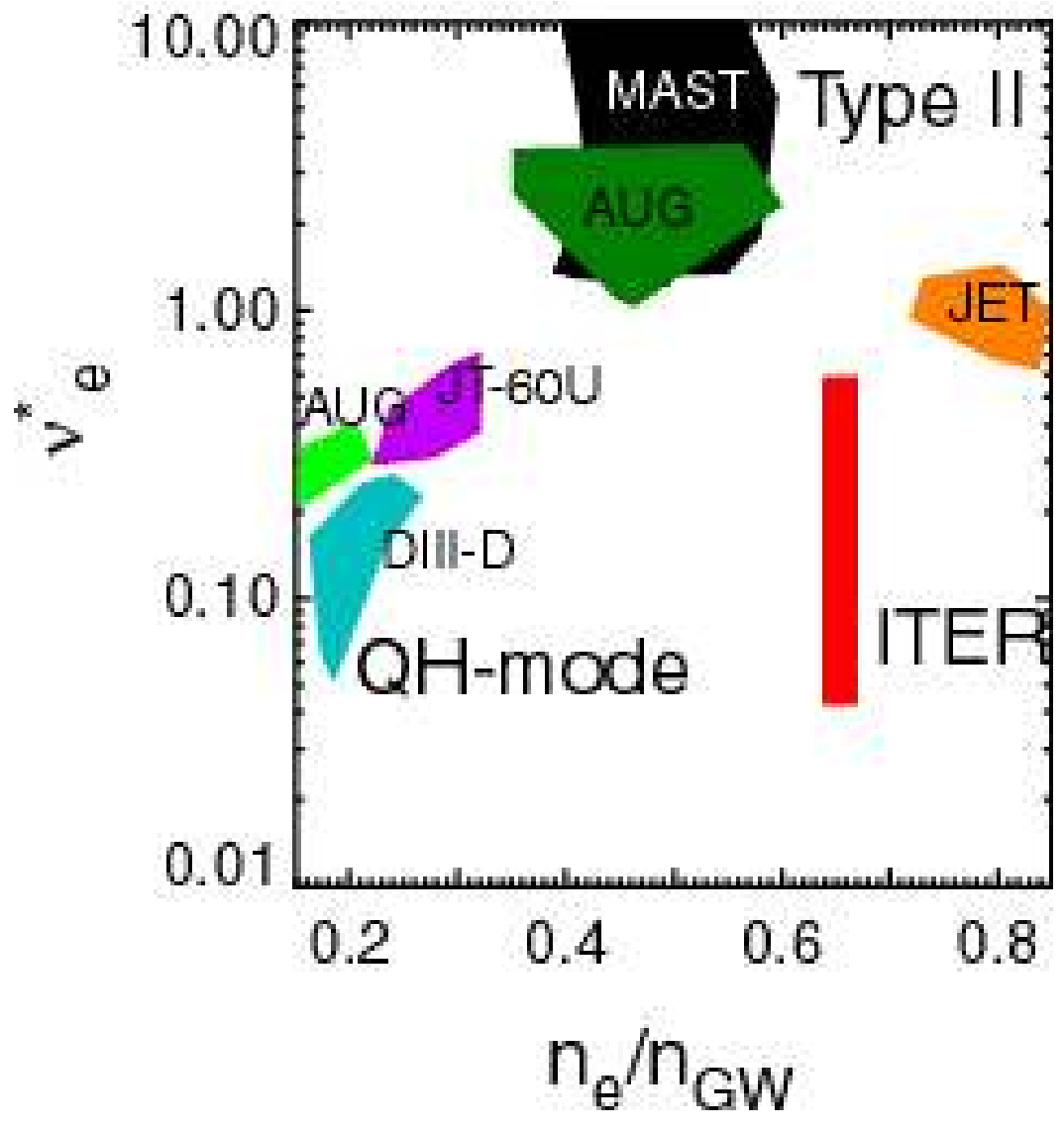

**Figure 8** Experimentally determined access condition in terms of pedestal collisionality ($\nu^*_e$) versus pedestal density as a fraction of the Greenwald density ($n_e/n_{GW}$) for type II ELMs and QH-mode.

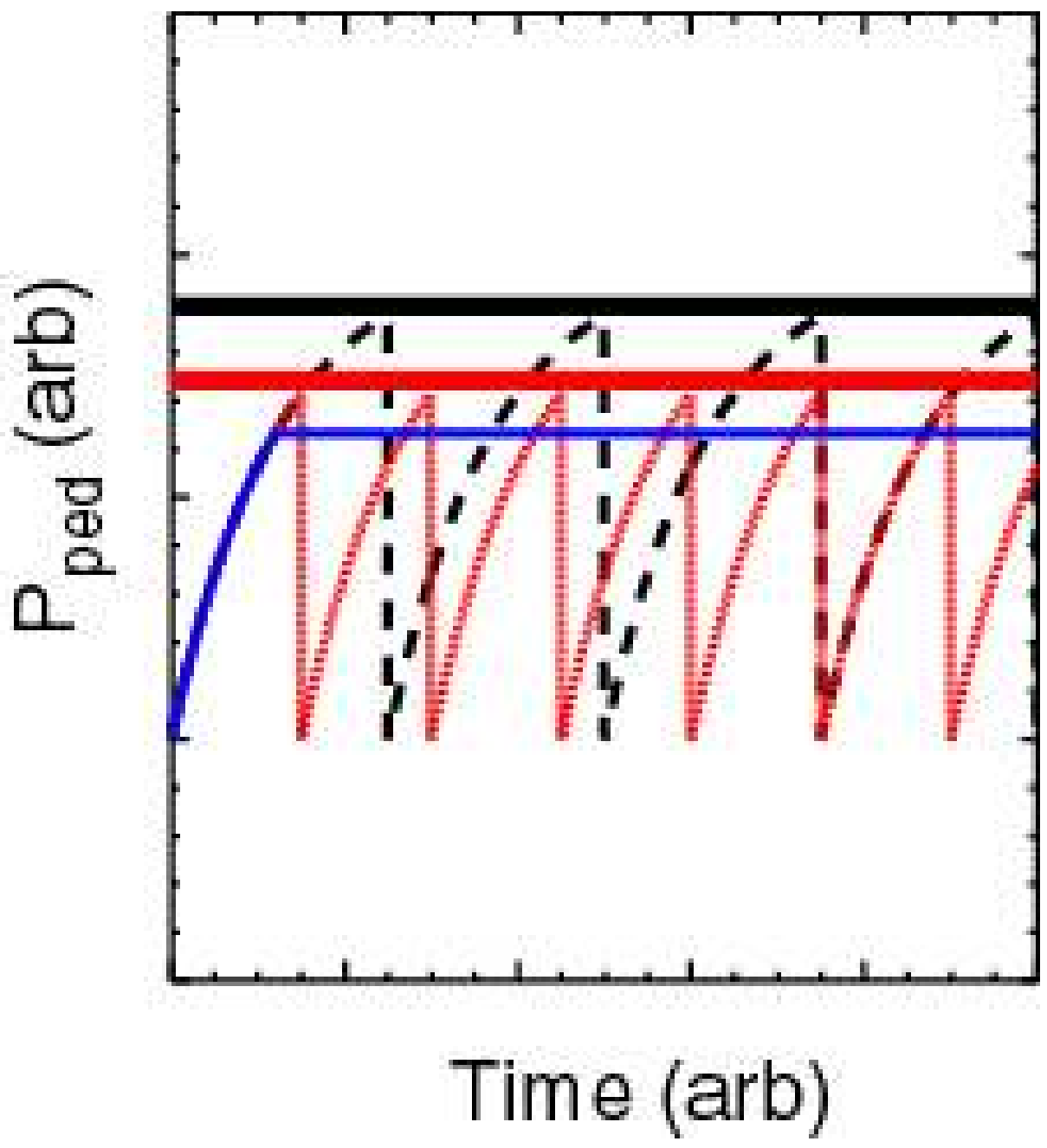

**Figure 9** Cartoon depicting the evolution of the pressure pedestal (curves) and the ballooning stability limit (horizontal line) during the ELM cycle for natural (dashed), mitigated (dotted) and ELM suppressed (solid).